\documentclass[twocolumn]{revtex4}
\usepackage{latexsym}
\usepackage{graphics,epsfig,graphicx}
\usepackage{epsfig}
\usepackage[german, english]{babel}
\usepackage[latin1]{inputenc}
\usepackage{color}
\usepackage{makeidx}

\begin{document}

\title{Phase-predictable tuning of single-frequency optical synthesizers}


\author{Felix Rohde,$^{1,*}$ Erik Benkler,$^1$ Thomas Puppe,$^2$ Reinhard Unterreitmayer,$^2$ Armin Zach,$^2$ and Harald R Telle$^1$}

\address{
$^1$Physikalisch-Technische Bundesanstalt, Bundesallee 100, Braunschweig D-38116, Germany, \\
$^2$TOPTICA Photonics AG, Lochhamer Schlag 19, D-82155 Graefelfing, Germany \\
$^*$Corresponding author: felix.rohde@ptb.de
}

\begin{abstract} We investigate the tuning behavior of a novel type of single-frequency optical synthesizers by phase comparison of the output signals of two identical devices. We achieve phase-stable and cycle slip free frequency tuning over 28.1 GHz with a maximum zero-to-peak phase deviation of 62 mrad. In contrast to previous implementations of single-frequency optical synthesizers, no comb line order switching is needed when tuned over more than one comb line spacing range of the employed frequency comb. 
\end{abstract}

\maketitle

\begin{figure}[tb]
\centerline{\includegraphics[width=\columnwidth]{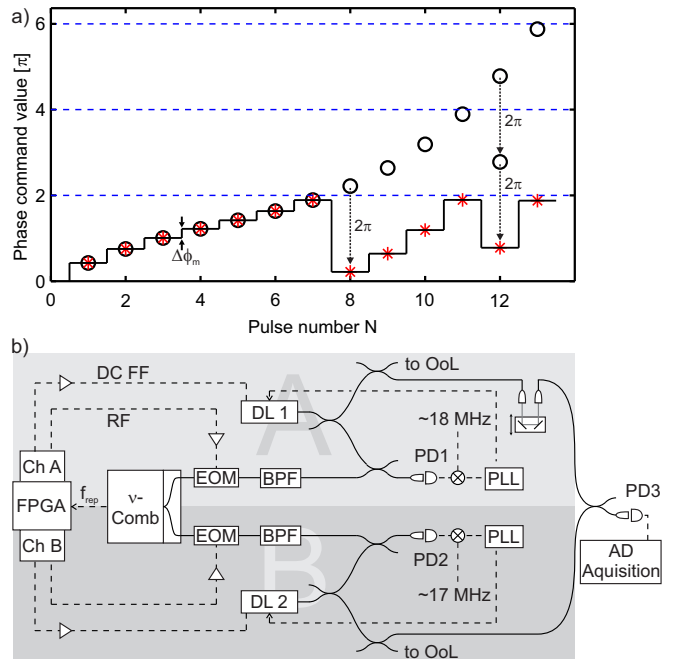}}
\caption{a: (Color online) Phase command values as a function of pulse number for a parabolic frequency evolution of the frequency comb. Circles: calculated values according to Eq.1, stars (red): values modulo $2\pi$, arrows: $2\pi$ flybacks, solid line: possible phase evolution at EOM. b: Setup of the experiment consisting of two channels ChA and ChB. FPGA: field programmable gate array, EOM: electro-optical modulator, BPF: band pass filter (50 GHz), DL1/2: external cavity diode laser 1 and 2, PD: photo diode, PLL: phase locked loop, OoL: out of loop measurement.}
\label{setup}
\end{figure}

\noindent A single-frequency optical synthesizer (SOS) is the analog of an electrical synthesizer in the optical domain. It provides a single-frequency optical field whose frequency and phase can be arbitrarily adjusted within a certain spectral range and resolution while it is related to a reference signal in a phase-coherent fashion. In an ideal case it combines the best of two worlds in a tunable way, i.e. the spectral resolution of narrow linewidth frequency stabilized lasers with the broad spectral coverage of frequency combs. \\
Such an SOS has applications in basic research e.g. in precision spectroscopy \cite{DelHaye2009}, frequency metrology and quantum optics but could also enable the development or improvement of a number of useful practical devices in optical metrology like high resolution optical spectrum or vector analyzers, phase stable cw THz-synthesizers \cite{Yasui2011}, high resolution, high dynamic range optical frequency domain reflectometers based on frequency scanning, optical coherence tomography \cite{Wieser2012} or traceable range finders and laser trackers.\\
The invention of frequency combs \cite{Holzwarth2000}, which consist of a broadband fixed frequency grid with well-defined frequency and phase relations between the individual comb lines, has enabled the most sophisticated SOS approaches up to date \cite{Jost2002,Schibli2005,Ahtee2009}. These implementations are based on phase-locking of a single-frequency ``clean-up'' laser to an individual comb line of a frequency comb and subsequent tuning of the repetition rate \cite{Ahtee2009} or tuning of the offset frequency between ``clean-up'' laser and comb line \cite{Jost2002,Schibli2005}. The tuning speed and range is hampered in the first approach due to the required macroscopic resonator length variation. \\
The approaches based on offset frequency tuning suffer from an ambiguity at ``critical'' optical frequencies, where the beat frequency between a specific comb line and the ``clean-up'' laser becomes zero or coincides with the beat frequency from an adjacent comb line. This leads to forbidden frequency gaps and limits the agility of the SOS. Moreover, these ambiguities need to be resolved in a complex way which imposes additional technical overhead. If forbidden frequency gaps at the critical frequencies are to be avoided, an offset frequency tuning based SOS needs to take a smart action when approaching these frequencies. In \cite{Schibli2005} for example, a second, frequency shifted beat note is generated and used as alternative error signal when critical frequencies are approached. In that case, the SOS will have problems to preserve phase stability when switching between error signals. \\
The aforementioned complications can be circumvented by combining a novel method for frequency shifting the carrier frequency of frequency combs that we have reported on in an earlier paper \cite{Benkler2013}, with a fixed offset phase lock of a ``clean-up'' laser to a single comb line. In this case, frequency tuning of the comb spectrum avoids any problematic ambiguity since all rf-beat frequencies are constant in time and conjugate beat signals never coincide. Such a universal SOS consists of three elements, i.e. a mode-locked laser acting as frequency comb generator (OFC), the mentioned frequency shifter for frequency combs \cite{Benkler2013} and a single line selector \cite{Rohde2013} which ensures single frequency operation. The latter is implemented with a phase lock of a ``clean-up'' laser to a single comb line. \\
In this paper we present the first characterization of the tuning properties of an SOS based on such a frequency shifter. In order to characterize the carrier phase fidelity of the SOS during tuning action we duplicate the system and use the second SOS as reference enabling the treatment in the rotating frame. We show that we can achieve phase-stable tuning of our SOS with respect to the reference over the full tuning range of the ``clean-up'' laser.\\
The principle of the frequency shifter relies on serrodyne frequency shifting \cite{Kohlhaas2012} the carrier frequency of an OFC. It exploits the fact that the spectrum of the OFC corresponds to a periodic pulse train in the time domain and shifts the optical carrier phase between subsequent pulses of the OFC. To explain the principle, the definition of the instantaneous frequency of the monochromatic signal of the comb line with order number $m$ can be used.
\begin{equation}
\nu_m(t)=\nu_{m,0}+\frac{1}{2\pi}\frac{\Delta\phi_{m}(t)}{\Delta T}.
\label{equ1}
\end{equation}
The frequency evolution $\nu_m(t)$ of the signal is given by a constant frequency $\nu_{m,0}$ plus a change of the carrier phase $\Delta\phi_{m}(t)$ per unit time $\Delta T$. In the frequency domain the frequency comb can be treated as a multitude of monochromatic fields, each experiencing the frequency evolution according to Eq.(\ref{equ1}). In the time domain the monochromatic fields interfere destructively at the dark intervals of the pulse train. The natural time interval for an incremental phase change is thus the temporal pulse-to-pulse spacing, i.e. $f_{\mathrm{rep}}^{-1}$. The pulse-to-pulse change $\Delta\phi_{m}(t)$ of the carrier phase is implemented by adding consecutively changing phase command values by means of an electro-optic phase modulator (EOM). The sequence of phase command values required for a target temporal evolution of the frequency shift is determined by temporal integration of the second term in Eq.(\ref{equ1}). A simple constant carrier frequency shift thus corresponds to a linear stepwise increase of the carrier phase between subsequent pulses. The cubic phase evolution of Fig.\ref{setup}a accordingly leads to a parabolic frequency evolution of the comb. Since the electromagnetic field is $2\pi$ periodic, the phase command values can be applied modulo $2\pi$, avoiding their divergence and thus of the voltages applied to the EOM. \\
The incremental phase changes $\Delta\phi_{m}(t)$ between pulses and the $2\pi$ flyback events in Fig.\ref{setup}a occur at the dark intervals of the pulse train. Due to the pulsed nature of the comb electromagnetic field, only the effective phase values which are present during the optical pulse length contribute to the frequency shifting effect. Thus, possible imperfections of the phase steps, e.g. transients or ringing, are concealed and the generation of spurious frequencies is suppressed. Consequently, the phase evolution between the pulses can be designed in an arbitrary fashion.\\
A detailed description of the frequency shifter for frequency combs and experimental proof that frequency tuning of a comb over multiple $f_{\mathrm{rep}}$ is readily achieved can be found in \cite{Benkler2013}.\\
In our experiment we use an integrated EOM (Eospace) to set the carrier phase of subsequent pulses of a frequency comb. 
The setup consists of two identical, but independent channels, A and B, as shown in Fig.\ref{setup}b. A fiber splitter divides the output power of a frequency comb generator (Toptica FemtoFErb SYNC, $\lambda$=1550nm, $f_{\mathrm{rep}}$= 56 MHz) in two equal parts of 30 mW average power and sends them to the two identical EOMs. The phase command values for both channels are generated with the help of two independent numerically controlled oscillators (NCO, details in \cite{Benkler2013}) that are implemented in one FPGA (Stratix III). The light fields in both channels are thus manipulated completely autonomously by digital-to-analog converting the phase command values and sending them to the EOM by means of an rf-signal. The FPGA is clocked with a harmonic of $f_{\mathrm{rep}}$, hence the changes $\Delta\phi_{m}(t)$ and the $2\pi$ flybacks of the phase command value are synchronized with the comb's pulse train. The signal gain of the phase command values is adjusted such that the $2\pi$ flybacks correspond to a carrier phase shift of $2\pi$ to the EOM. For the sake of simplicity we describe only channel A in the following. The transmitted, frequency shifted light is filtered by a 50 GHz band-pass filter to improve the signal to noise ratio of the beat between a selected comb line $m$ and an external cavity diode laser (DL1, Toptica DL pro design, $\lambda$=1550nm). The light fields from DL1 and the filtered comb are superimposed on a photodiode (PD1) using polarization maintaining fiber splitters. The electric beat signal of the PD is used to phase-lock DL1 to the selected comb line $m$. The feedback loop is closed using a fast analog controller (Toptica FALC 110) which acts on the piezo and current of DL1. Additionally, the FPGA presets the desired frequency evolution of DL1 by means of a baseband feed-forward signal to the piezo of the laser. Thus, the phase locked loop (PLL) has to compensate only for imperfections of the feed-forward control, e.g. due to piezo hysteresis. \\
The remaining light from DL1 is split in two parts by a polarization maintaining fiber splitter. One part is sent to an out-of-loop measurement (OoL) of the laser frequency implemented with a delayed self-heterodyne interferometer of known length. The other part is sent to a photodiode (PD3) where it is superimposed with the corresponding light field from channel B, i.e. the light field from DL2 phase-locked to the same comb line $m$. The offset frequencies of the phase locks of laser DL1 and DL2 to the selected comb line $m$ differ by 1 MHz. The resulting 1 MHz beat signal is detected at PD3, digitized with a sampling rate of 7 MHz by a 16 bit DAQ card and stored on a computer. In a post-processing step, the differential phase of the two light fields during a desired frequency scan is extracted. This phase deviation is deduced from the raw data using a software phase-retrieval algorithm \cite{Servin1993}. Effectively, it compares the phase of the 1 MHz beat with the differential phase between the PLL offset signals. Since both channels share the same frequency comb generator, frequency and phase fluctuations of this ``master oscillator'' are ruled out via common-mode rejection and solely the phase deviation introduced by the frequency shifter is measured. \\ 
The setup in Fig.\ref{setup}b represents a two-beam interferometer with two separate paths from the EOMs to PD3. Both signal path lengths of about 11 m have to be matched to a degree of 300 $\mathrm{\mu m}$ in order to avoid moving spectral fringes during tuning. Length matching is accomplished with the help of a free space delay line in channel A. \\
To characterize the tuning behavior of the SOS both lasers are swept simultaneously over 500 comb lines, or 28.1 GHz following a sinusoidal frequency evolution. 
\begin{figure}[tb]
\centerline{\includegraphics[width=\columnwidth]{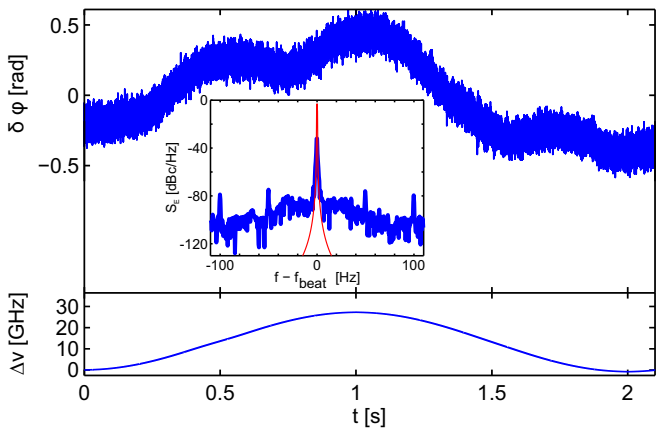}}
\caption{Top: Relative phase between DL1 and DL2 during a sinusoidal frequency sweep over 500 comb lines (28.1 GHz). Inset: (Color online) Power spectral density $S_{\mathrm{E}}$ of the beat note between DL1 and DL2 during the same sweep (blue). Sidebands at 50 and 100 Hz are caused by power line noise. FFT-filter function used to calculate $S_{\mathrm{E}}$ (red). Bottom: Result of the out-of-loop frequency measurement of DL1 during the sweep. The coarse frequency measurement has an uncertainty of $\pm 100$ MHz.}
\label{fig3}
\end{figure}
Fig.\ref{fig3} shows the phase deviation between DL1 and DL2 (top) and the frequency of DL1 (bottom) for this experiment. During this 2 s long sweep the relative phase is stable within $\pm 0.5$ rad. High frequency technical noise contributions ($\Delta\phi_{\mathrm{rms}}$ = 55 mrad, 4 Hz $< f <$ 3.5 MHz) are found together with a slow drift on the timescale of seconds. This slow drift is caused by fluctuations of the interferometer path length difference which is not controlled. Although the frequency shifting method is in principle capable of controlling the absolute phase value of the optical output field in a deterministic way, this has not yet been implemented here. Unwanted interferometric phase drifts could be reduced, e.g. by reducing the length of the fiber pigtails of the components, temperature stabilizing the interferometer setup or ultimately by active fiber length stabilization. \\
The inset in Fig.\ref{fig3} shows in blue the power spectral density $S_{\mathrm{E}}$ (rf-spectrum) of the beat note between both DLs during their common, 28.1 GHz wide frequency sweep. $S_{\mathrm{E}}$ was calculated from the raw data of Fig.\ref{fig3} (top) using FFT. Its 3dB-width of 0.7 Hz is slightly broader than the FFT-filter function (in red, 0.5 Hz) used to calculate the spectrum and the peak is 30 dB suppressed with respect to the filter function. This reflects the onset of a carrier collapse of the filtered rf-signal and is in accordance with the interferometric phase drifts found in Fig.\ref{fig3} (top). The spectrum represents a beat-note in the ``rotating-frame'' and is characteristic for the two-SOS-scheme employed here.\\
The tuning speed of the SOS reaches 45 GHz/s at the steepest slope of the 28.1 GHz sinusoidal modulation or up to 120 GHz/s for frequency modulations on the order of 200 MHz. Tuning speed and range are solely limited by the ``clean-up'' technique, i.e. by the bandwidth of the locking electronics and the mode-hop free tuning range of the diode laser. The frequency shifting method itself, disregarding the clean-up laser, is capable of tuning speeds of up to tens of THz per second combined with a tuning range over the whole bandwidth of the frequency comb. \\
To illustrate that the SOS is running cycle slip free, we intentionally provoke a cycle slip in the PLL of DL1 by a modification of the locking parameters. 
\begin{figure}[tb]
\centerline{\includegraphics[width=\columnwidth]{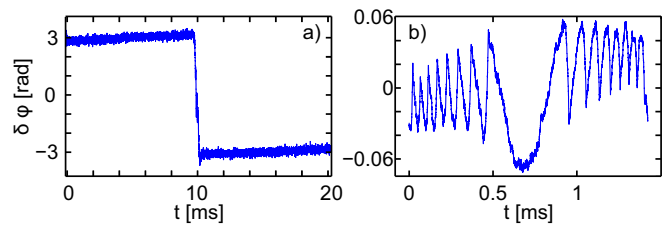}}
\caption{a: Relative phase between DL1 and DL2 during an experiment where a cycle slip has been provoked in the PLL of DL1. b: Average of 100 consecutive sweeps of the relative phase between DL1 and DL2 when DL1 is swept over a critical frequency (unshifted comb line).}
\label{fig4}
\end{figure}
The cycle slip manifests itself as a $2\pi$ jump in the differential phase between DL1 and DL2 as shown in Fig.\ref{fig4}a. From the absence of this signature in measurements with optimized locking parameters, we deduce that the SOS is operating cycle slip free.\\
In the following an experiment to characterize the phase deviations at the critical frequencies mentioned in the introduction is described. For our SOS the most relevant critical frequencies correspond to the frequencies of the initially unshifted comb lines. This can be explained  by small deviations of the EOM driving signal from the ideal shape, e.g. non perfect-$2\pi$-flybacks, which are sampled by the light pulses and lead to spurious sidebands at the optical output spectrum of the frequency shifter. The spurs occur at the frequencies of the initial unshifted comb lines and are typically suppressed by 30 dB with respect to the shifted comb lines. These spurs in the frequency domain induce PLL error signal glitches upon tuning the ``clean-up'' laser across the spur and show up as systematic phase deviations with amplitudes comparable to the technical noise of Fig. \ref{fig3} in our measurements. To measure the systematic phase deviations of an individual SOS, any common-mode suppression of the glitch during tuning transits across the spur has to be avoided. DL1 and DL2 are thus linearly swept over the spur with a frequency shifter induced detuning of 5 MHz. Then, phase-tracking their beat note while tuning reveals the individual phase deviations at the glitch positions since the individual transits of both SOS take place at different time instants. Thus, the true phase deviation of DL1 at the critical frequency is measured, while DL2 acts as an unperturbed local oscillator. Fig.\ref{fig4}b shows the phase deviation produced when DL1 is slowly swept over such an unshifted comb line. The trace shows the average of 100 consecutive sweeps with interferometric phase offsets subtracted. The averaging reduces the technical noise and reveals a systematic phase glitch with a zero-to-peak deviation of 62 mrad. The interference between the swept signal and the spur leads to an oscillatory shape of the glitch. The deviation from a pure oscillatory shape results from the PLL response. The observed phase glitch is consistent with the expected phase error caused by spurs which are 30 dB below the carrier. To the best of our knowledge this is the first time phase stable tuning of an SOS across critical frequencies is achieved and characterized. \\
The frequency resolution of the frequency shifter is given by the product of $f_{\mathrm{rep}}$= 56 MHz and the length of the NCO tuning word. The algorithm implemented in the FPGA firmware uses a phase resolution of $2\pi/48$ bits, leading to a frequency resolution of the SOSs center frequency of 200 nHz. It should be noted that the smaller amplitude resolution (12 bit) of the digital to analog converter used to generate the analog phase command values does not influence the frequency resolution, but determines the so called spur-free dynamic range and thus the maximum phase error. The frequency resolution could even be improved by using more bits in the FPGA implementation. \\
The short-term linewidth of DL1 locked to the free running unshifted comb corresponds to the linewidth of DL1 in the non-rotating frame and was measured to be well below 10 kHz at 1550 nm using a delayed self-heterodyne interferometer. The minimum absolute frequency uncertainty achievable with this SOS depends on the choice of frequency reference for the frequency comb and can approach the values of the best optical/microwave clocks.\\
In conclusion, we have demonstrated phase-predictable tuning of two independent, common mode suppressed SOSs. In contrast to previous implementations of SOSs, our approach allows for arbitrary, set-point-switching free tuning over the full spectrum of the used comb oscillator, limited only by the applied ``clean-up'' technique (28.1 GHz). The maximum zero-to-peak deviation of the phase found at critical frequencies was shown to be 62 mrad and the maximum tuning speed realized was 120 GHz/s. In combination with well studied methods for self-referencing of frequency combs and widely tunable lasers, this technique could stimulate the development of a compact SOS, capable of delivering a tunable absolute optical frequency of high precision. \\

\section*{Acknowledgments}
We gratefully acknowledge project management support by P. Leisching. The work was funded by the German Ministry of Economics and Technology (ZIM projects KF2303709 and KF2806204).


\end{document}